\documentstyle[prl,aps,epsfig]{revtex}

\begin{document}
\def\be{\begin{equation}}
\def\ee{\end{equation}}
\def\bea{\begin{eqnarray}}
\def\eea{\end{eqnarray}}
\def\bml{\begin{mathletters}}
\def\eml{\end{mathletters}}
\def\l{\label}
\def\b{\bullet}
\def\eqn#1{(~\ref{eq:#1}~)}
\def\no{\nonumber}
\def\av#1{{\langle  #1 \rangle}}

\title{Evolutionary trajectories in rugged fitness landscapes}

\author{Kavita Jain and Joachim Krug}
\address{Institut 
f\"ur Theoretische Physik, Universit\"at zu K\"oln, Z\"ulpicher Strasse 77, 
50937 K\"oln, Germany}
\maketitle
\widetext

\begin{abstract}
We consider the evolutionary trajectories traced out by an infinite population
undergoing mutation-selection dynamics in static, uncorrelated random
fitness landscapes. Starting from the population that consists of a single 
genotype, the most populated genotype \textit{jumps} from a 
local fitness maximum to another and eventually reaches the global maximum. 
We use a strong selection limit, which reduces the dynamics beyond the 
first time step to the competition between independent mutant subpopulations,  
to study the dynamics of this model and of a simpler one-dimensional model 
which ignores the geometry of the sequence space. 
We find that the fit genotypes that appear along a trajectory are a subset
of suitably defined fitness \textit{records}, and exploit several results
from the record theory for non-identically distributed random variables.
The genotypes that contribute to the trajectory are those records that are
not \textit{bypassed} by superior records arising further away from the
initial population. Several conjectures concerning the statistics of 
bypassing are extracted from numerical simulations. In particular, for the 
one-dimensional model, we propose a simple relation between the bypassing 
probability and the dynamic exponent which describes the scaling of the 
typical evolution time with genome size. The latter can
be determined exactly in terms of the extremal properties of the
fitness distribution. 
\vskip0.5cm
\noindent PACS numbers: {87.10.+e, 87.23.Kg, 05.40.-a}
\end{abstract}

\section{Introduction}
\l{intro}

The episodic nature of biological evolution has provided motivation for much
work on the modeling of evolutionary dynamics in the statistical physics
community \cite{Sneppen95,Sibani98}; see 
\cite{Peliti97,Baake99,Drossel01} for review. Evolution displays
a \textit{punctuated} pattern, with epochs of no or slow change 
interspersed with
bursts of (relatively) rapid activity, on various levels ranging from
the fossil record \cite{Eldredge89,Gould93} to experiments with microbial 
populations \cite{Lenski94,Elena96,Burch99,Elena03}. Punctuated behavior is 
also seen in simulations of \textit{in vitro} evolution
of RNA molecules \cite{Schuster02}, optimization algorithms 
\cite{Rujan88,vanNimwegen97}, and artificial life \cite{Adami95}. 
It has been recognized for a long time that one possible 
scenario that is consistent with punctuated dynamics is the evolution 
of a population in a static fitness landscape with many peaks.
In this picture, the two modes of evolution correspond to the extended
periods of residence of the population at a local fitness maximum
and the exploration of a new, higher lying peak, respectively,
which entails the rapid crossing of a valley of lower fitness 
\cite{Newman85,Lande85}. 
In essence, this is the phenomenon of quantum evolution described by 
the paleontologist G.G. Simpson sixty
years ago \cite{Simpson44}, and more recently referred to in 
macroevolutionary theory 
as punctuated gradualism \cite{Eldredge89}. 
If the population is large, so that the distribution of individuals over the 
various phenotypes or 
genotypes can be modeled by a continuous field, the transition between 
fitness peaks described above displays analogies with physical processes  
such as quantum tunneling \cite{Ebeling84}, variable range hopping 
\cite{Zhang86} or noise-driven barrier crossing. 

Since this metastable behavior seems ubiquitous in nature, it may be 
worthwhile to study a simple model where this can be 
analysed in detail. A convenient mathematical framework 
to address this issue is the quasispecies model which was originally 
introduced to describe large populations of self-replicating 
macromolecules \cite{Eigen71,Eigen89}. The quasispecies model is a 
mutation-selection model whose steady state has been studied in great detail.  
For various choices of fitness landscapes, it exhibits 
the phenomenon of error threshold in which beyond a 
critical mutation rate, the population delocalises over the whole sequence  
space. 

In this paper, we address the question of punctuated \textit{dynamics} in the 
quasispecies model. To avoid complications associated
with the error threshold phenomenon \cite{Baake99,Eigen89}, we work   
in a strong selection limit inspired by the zero temperature limit
of the statistical mechanics of disordered systems \cite{Krug02} 
(for a different kind of strong selection
limit used in population genetics see e.g. \cite{Woodcock96}). 
In this limit, the location of the population in the space of genotypes
can be identified with the most populated genotype at all times, 
and the evolutionary trajectories can be represented in a particularly
transparent manner \cite{Krug03}. 
Since the population is located at a single genotype at any time, the 
evolutionary trajectory changes in a stepwise manner. To generate an 
evolutionary 
trajectory, a localized population is placed at a randomly chosen point 
in the space of all possible genotypes. Due to infinite population, in the 
next time step, all genotypes get occupied with nonzero population. 
As we describe later in detail, it turns out that 
the fit genotypes typically receive small initial 
population. Strong selection then reduces the problem to competition 
between these fit subpopulations struggling to overcome the poor 
initial conditions.

In the next section, we describe
the quasispecies model and derive the reduced representation 
that arises in the strong selection limit. The bulk of the 
paper is then devoted to the analysis of the strong selection dynamics, which
turns out to be rather closely related to the mathematical theory of records
\cite{Glick78,Nevzorov87,Arnold98,Nevzorov01}; a relation between 
biological evolution and record statistics has been proposed 
previously \cite{Sibani98,Kauffman87}, see also \cite{Krug04}. 

\section{Quasispecies dynamics in the strong selection limit}
\l{model}

The quasispecies model is defined on the space of genotypes represented by 
sequences $\sigma \equiv \{ \sigma_{1},...,\sigma_{N} \} $,  
where each of the $N$ letters $\sigma_{i}$ is taken from an alphabet of size
$\ell \geq 2$. The number of individuals of genotype $\sigma$ 
at time $t$ is represented
by a real variable $Z(\sigma,t)$ that obeys the discrete time 
evolution equation 
\be
Z(\sigma,t+1)= \sum_{{\sigma}^{\prime}}
p(\sigma^{\prime} \rightarrow \sigma) \; W(\sigma^{\prime}) 
\; Z(\sigma^{\prime},t).
\l{full}
\ee
Here the fitness $W(\sigma)$ is defined \cite{Peliti97} as 
the expected number of offspring produced by an 
individual carrying sequence $\sigma$, 
and $p(\sigma^{\prime} \rightarrow \sigma)$ is the probability that
genotype $\sigma$ is created as offspring of genotype $\sigma'$ due to 
copying errors in the genome. A simple choice for the latter 
corresponds to independent point mutations occurring with 
probability $\mu$ per generation, so that 
$p(\sigma^{\prime} \rightarrow \sigma)= 
(\mu/(\ell-1))^{d(\sigma^{\prime},\sigma)} 
(1-\mu)^{N-d(\sigma^{\prime},\sigma)}$,  
where 
\be
d(\sigma^{\prime},\sigma) = \sum_{i=1}^N (1 - \delta_{\sigma_i,\sigma_i'})
\;\;, \l{Hamming}
\ee
is the Hamming distance between sequences $\sigma^{\prime}$ and $\sigma$.

A constraint of fixed population size could be enforced by dividing
the right hand side of (\ref{full}) by the population averaged fitness
$\langle W \rangle$. This introduces an (inessential)
nonlinearity into the problem \cite{Peliti97}, which is why we prefer to work 
with the
unnormalized population variables $Z(\sigma,t)$.   
It is clear that within the quasispecies framework, which requires an
infinite population from the outset, the actual population size cannot
play any role.

To derive the strong selection limit \cite{Krug02}, we 
write $Z(\sigma,t)=e^{\kappa E(\sigma,t)}$, 
$W(\sigma)=e^{\kappa F(\sigma)}$ 
and $\mu=e^{-\kappa}$ where $\kappa$ is the inverse selective 
temperature \cite{Peliti97,Franz97} 
and $E(\sigma,t)$ and $F(\sigma)$ are logarithmic population and
fitness variables, respectively. Throughout this paper we take
the fitness landscape to be completely uncorrelated, which implies that
the fitnesses $F(\sigma)$ are independent, identically 
distributed (i.i.d.) quenched random variables chosen 
from a distribution $p(F)$
with support on interval $[F_{\mathrm{min}},F_{\mathrm{max}}]$. 
The strong selection limit corresponds to $\kappa \rightarrow \infty$, 
which yields
\be
E(\sigma, t+1)= \mbox{max}_{\sigma^\prime} \left[ E(\sigma^\prime,t)+ 
F(\sigma^\prime)-d(\sigma,\sigma^\prime) \right] \l{fullss} \;\;.
\ee
Starting with an initial condition 
$Z(\sigma,0)=\delta_{\sigma,\sigma^{(0)}}$ in which only a 
single, randomly chosen  
sequence $\sigma^{(0)}$ has nonzero population, at the next time step each 
sequence gets seeded with the logarithmic population
\be
\l{sslimit2}
E(\sigma,1) = F(\sigma^{(0)}) - d(\sigma,\sigma^{(0)}).
\ee
In terms of the original variables $Z(\sigma,t)$, this
corresponds to a population distribution that decays exponentially
with increasing Hamming distance from $\sigma^{(0)}$. 
Remarkably, it turns out that for the subsequent time evolution the 
mutations are unimportant for genotypes with 
high fitness. Since we are primarily interested in such genotypes, the 
dynamics of the model can be approximated by allowing each genotype 
to reproduce with its intrinsic rate $F(\sigma)$ for $t > 1$. 
In terms of the logarithmic variables, this leads to
\be
E(\sigma,t) = E(\sigma,1)+(t-1) F(\sigma). \l{sslimit1} 
\ee
Thus, (\ref{fullss}) reduces to a problem of non-interacting sequences whose 
population is growing linearly in time. 
The approximation of ignoring mutations after the seeding 
stage was tested numerically \cite{Krug03} and found to be in 
very good agreement with the full model described by (\ref{fullss}). 

A further simplification can be made by noting that 
due to (\ref{sslimit2}), the population $E(\sigma,1)$ 
is same for all the sequences that lie 
on a shell of constant Hamming distance $d(\sigma^{(0)},\sigma)$ away from the 
sequence $\sigma^{(0)}$. Since we will be interested in the sequence with 
the largest 
population at any instant, only the sequence with the largest 
fitness within a shell needs to be considered. Labeling a shell by its 
Hamming distance $k=0,1,...,N$ from the  sequence $\sigma^{(0)}$, we arrive 
at the {\it shell model} \cite{Krug03}
in which the population $E(k,t)$ of the fittest sequence 
in the shell $k$ with a total number of $\alpha_{k}$ sequences obeys 
\be
\l{lineq}
E(k,t)=-k + F(k) t \l{shell}.
\ee
Here $F(k)$ is the maximum of $\alpha_k$ 
i.i.d. random variables drawn from the fitness
distribution $p(F)$; hence the $F(k)$ are independent 
but non-identically distributed random variables. 
To arrive at the simple form (\ref{shell}), we have redefined the 
population as $E(k,t)+F(k)-F(0)$. 
The number of sequences in shell $k$ is given by the expression 
\be
\l{alphak}
\alpha_{k}= {N \choose k} \; (\ell-1)^{k},
\ee
as can be seen by noting that 
there are ${N \choose k}$ ways of choosing $k$ letters at which a 
sequence $\sigma$ differs from $\sigma^{(0)}$, and each 
of these $k$ letters can take $\ell-1$ values.
For large $N$, the majority of sequences is contained in a belt of width 
$\sim \sqrt{N}$
around the distance $k_{\mathrm{max}} = N (\ell -1)/\ell$ 
where (\ref{alphak}) is peaked. 

We will also consider an {\it i.i.d. model} for which the population $E(k,t)$ 
evolves according to (\ref{shell}) but the fitnesses $F(k)$ are i.i.d. 
random variables. This choice corresponds to a one-dimensional sequence space 
and we will see that our results are sensitive to the geometry of the 
sequence space. The i.i.d. model is closely related to the 
zero temperature limit of the problem of
directed polymers in the presence of columnar defects studied in 
\cite{Krug93} and a version of the parabolic
Anderson model \cite{Gaertner04}.   

\begin{figure}
\begin{center}
\psfig{figure=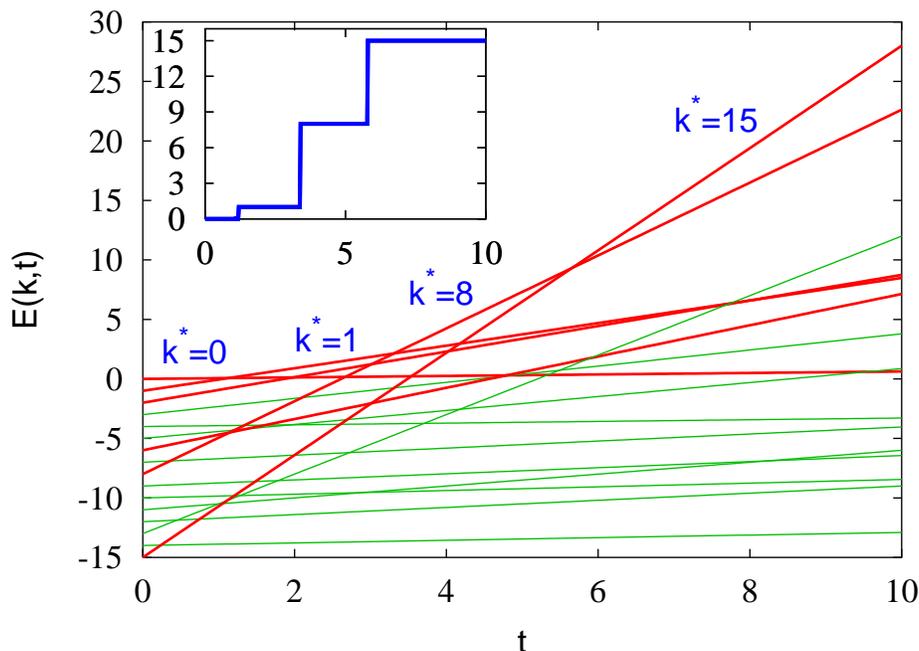,width=9cm,angle=270}
\caption{
Time evolution of the logarithmic population variable $E(k,t) = - k +
F(k)t$. The fitnesses $F(k)$ of the sequences corresponding to thin green
lines are not records whereas those corresponding to bold red lines
are. The lines appearing in the upper envelope define the most populated
sequence $k^\ast$; an evolutionary jump occurs when two such lines cross.
Note that the sequences $k=2$ and $k=6$, although constituting fitness
records, are bypassed as they do not satisfy the minimum crossing time
constraint. The inset shows the punctuated evolution of the most
populated sequence $k^\ast$ as a function of time.}
\label{lines}
\end{center}
\end{figure}

Figure \ref{lines} illustrates the geometric picture of the evolutionary
process (\ref{shell}) 
that emerges after these simplifications and approximations. 
At a given time $t$, the most populated genotype
$k^\ast$ for which $E(k^\ast,t) = \max_k \{ E(k,t) \}$ leads until it is 
overtaken by a sequence $k'^\ast$ and so on.  
While the last leader is obviously located at the global fitness maximum, the 
identity of the previous leaders is non-deterministic. 
The inset of 
Fig.~\ref{lines} shows the punctuated evolution of the leading 
sequence $k^{\ast}$. 
Our main interest in the present work is in the statistical properties
of these leadership changes or evolutionary jumps. 
Initially, the sequence $\sigma^{(0)}$ with population $E(0,t)$ leads.
A shell $k'$ can overtake the currently leading shell 
$k < k'$ at the crossing time $T({k,k^{\prime}})$ given by 
\be
T(k',k)= \frac{k'- k} {F(k')-F(k)} \;\;, \l{Tkkprime}
\ee
which is positive provided $F(k') > F(k)$. In 
Fig.~\ref{lines}, only the shells $k=1, 2, 6, 8, 15$ 
can overtake $k=0$ (and only $k=2, 6, 8, 15$ can overtake $k=1$, ...)
since $F(15) > F(8) > F(6) > F(2) > F(1) > F(0)$. Such a set of fitness in 
which each value is progressively 
larger than all the previous ones defines a sequence of {\it{records}}. 
However, in order 
to appear in the evolutionary trajectory, it is not sufficient to be a record. 
As shown in Fig.~\ref{lines}, the shell $k=2$ is 
bypassed by shell $k=8$ since 
$T(8,1)=\mbox{min}_{k > 1} T(k,1)$. In general, if the current 
leader is in shell $k$, then the next leader is in the shell $k'$ for which 
the crossing time (\ref{Tkkprime}) is \textit{minimized} 
\cite{Rujan88}. Thus, in principle, the properties of leadership
changes can be formulated in terms of the extremal statistics of the 
matrix of crossing times. However, as will be explained in more detail in 
Sect.\ref{mft}, this minimization problem is too cumbersome 
to handle analytically and is further complicated by the presence of 
strong correlations among the crossing times.    

The nontrivial dynamics of the change in leadership described above is 
due to the competition between the initial population at a sequence and its 
fitness. As discussed above, a potential leader must be a record and hence 
must occur at a large Hamming distance away from the sequence $\sigma^{(0)}$
but the initial population at such a sequence is small 
due to (\ref{sslimit2}). Thus, the disadvantage due to poor initial condition 
may be overcome by a better fitness. Further, since 
the fitness of the leader is correlated to its Hamming distance from the 
sequence $\sigma^{(0)}$, the fitness $F(k^*)$ also shows punctuated evolution.
 
The rest of the paper is organised as follows.
In Sections~\ref{record} and \ref{jump}, we define records and jumps precisely 
and study some of their statistical properties, both for the shell model
and for the i.i.d. model. In Section~\ref{approach}, we 
describe the dynamics of the approach to the global maximum of the fitness 
landscape. A conjecture
concerning the probability of bypassing in the i.i.d. model, 
based on the results obtained in Sections~\ref{jump} and \ref{approach} 
along with numerical evidence, is presented in Section~\ref{beta-z}.
In Section~\ref{mft}, we introduce and discuss a further simplification of the 
problem described by (\ref{shell}), which highlights the strong 
interdependence of the crossing times.  
Finally, we summarise our results and conclude 
with a brief discussion of the possible biological relevance of this work 
in Section~\ref{conclude}.

\section{Record statistics in sequence space}
\l{record}

In a sequence $\{ X_{k} \}$ of random variables, an upper (or lower) 
record is said to occur at $m$ if $X_{m} > X_{k}$ (or $X_{m} < X_{k}$) 
for all $k < m$ \cite{Glick78,Nevzorov87,Arnold98,Nevzorov01}. Since only the upper 
records are pertinent to our study, we shall henceforth refer to them as 
records. In the following subsections, we study some characteristics of 
these records; in particular, we find the mean number of records and 
the typical 
spacing between them. The record statistics of i.i.d. variables is 
well studied 
and we briefly review some of the known results as they are useful for 
later discussion. We study the record statistics in the shell model, for which 
the random variables are not identically distributed, in some detail. 
An appealing general feature of record statistics is that the results
are independent of the underlying probability distribution $p(F)$, 
which therefore does not need to be specified in this section. 

\subsection{Number of records}

In this subsection, we calculate the average number 
${\cal{R}}_{\mathrm{shell}}$ of 
records in the shell model. 
For i.i.d. random variables, it is well known that the average number of records
among $N$ variables is ${\cal{R}}_{\mathrm{iid}} \approx \ln N$ for large $N$. To see 
this, note that the probability 
$\tilde{P}_{\mathrm{iid}}(k)$ that the $k$'th random variable is a record is equal to the 
probability 
that it is the largest among the first $k$ random variables. 
Since the location of the global 
maximum is uniformly distributed for i.i.d. variables, it follows that 
$\tilde{P}_{\mathrm{iid}}(k)=1/k$ for any choice of $p(F)$. 
Summing $\tilde{P}_{\mathrm{iid}}(k)$ up to $k=N$ yields the average number 
${\cal{R}}_{\mathrm{iid}}$ of records to be $\ln N$. In fact, it 
is known that the probability distribution of the number of records is 
a Poisson distribution with mean $\ln N$ \cite{Sibani98}. 

The record statistics for non-i.i.d. variables is much less studied. 
A class of models in which records are obtained from a sequence $\{ X_{k} \}$ 
where each $X_{k}$ itself is the maximum of a set of $\alpha_{k}$ 
i.i.d. random variables was considered by Nevzorov \cite{Nevzorov84}. 
Such models have been used, for instance, in an (unsuccessful) attempt to 
account for the frequent occurrence of Olympic records due to an 
increasing world 
population \cite{Yang75}. The i.i.d. model is a special case  
corresponding to $\alpha_{k}=1$ for all $k$ and 
the shell model is obtained by using $\alpha_k$ given by (\ref{alphak}).

To analyze the Nevzorov model, 
we define a binary random variable $Y_{k}$ which takes value $1$ 
if a record occurs in the $k$'th set and $0$ otherwise. Since the distribution 
$p_{k}(F)$ of the maximum of $\alpha_{k}$ i.i.d. random variables is given by \cite{David70}
\be
\l{pk}
p_{k}(F)= \alpha_{k} \; p(F) \; \left( \int^{F}_{F_{\mathrm{min}}} p(x) dx \right)^{\alpha_{k}-1},
\ee 
we have
\be
\mbox{Prob}(Y_{k}=1)=\int_{F_{\mathrm{min}}}^{F_{\mathrm{max}}} dF \; 
\prod_{i=0}^{k-1} q_{i}(F) \;\; p_{k}(F)= 
\frac{\alpha_{k}}{(\alpha_{0}+...+\alpha_{k})} \;\;,\l{PYk1}
\ee
where $q_{k}(F)$ is the cumulative distribution corresponding to $p_{k}(F)$. 
The above equation expresses the fact that the $k$'th record value is a 
global maximum amongst $\sum_{j=0}^{k} \alpha_{j}$ variables 
and there are $\alpha_{k}$ ways in which it can occur in the $k$'th set. 
Further, the joint probability $\mbox{Prob}(Y_{k_{1}}=1,Y_{k_{2}}=1)$ for 
$k_{1} < k_{2}$ is given by 
\be
\mbox{Prob}(Y_{k_{1}}=1,Y_{k_{2}}=1)= \int_{F_{\mathrm{min}}}^{F_{\mathrm{max}}} dF \; 
\prod_{i=0}^{k_{1}-1} q_{i}(F) \;\;  p_{k_{1}}(F)
\int_{F}^{F_{\mathrm{max}}} dG \;\prod_{j=k_{1}+1}^{k_{2}-1} q_{j}(G) \;\; 
p_{k_{2}}(G)= 
\frac{\alpha_{k_{1}} \; \alpha_{k_{2}}}{(\alpha_{0}+...+\alpha_{k_{1}}) \;
(\alpha_{0}+...+\alpha_{k_{2}})} \;\;. \l{PYk12}
\ee
In a similar manner, it can be shown that 
$\mbox{Prob}(Y_{k_{1}}=1,...,Y_{k_{m}}=1) =
\prod_{j=1}^{m} \mbox{Prob}(Y_{k_{j}}=1)$ for any $m \ge 2$. 
Thus, the $Y_{k}$'s are 
independent, non-identically distributed variables 
\cite{Nevzorov87,Arnold98,Nevzorov01}. 

For the shell model, due to (\ref{alphak}) and (\ref{PYk1}), the probability 
$\tilde{P}_{\mathrm{shell}}(k,N)$ that a record occurs in the $k$'th shell is given by 
\be
\tilde{P}_{\mathrm{shell}}(k,N)=\mbox{Prob}(Y_{k}=1) \approx 1- \left( \frac{1}{\ell-1} \right) 
\frac{a}{1-a} \;\;\;\;,\;\;\;\; a= \frac{k}{N} < \frac{\ell-1}{\ell} =
\frac{k_{\mathrm{max}}}{N} \;\;,
\l{Ptildeshell}
\ee 
where we have used that 
${N \choose k-m} / {N \choose k} \approx [a/(1-a)]^{m}$ for $k, N \gg 1$ with 
$k/N$ fixed. Since it is easier to break records in the beginning, the 
probability to find a record is near unity for
$k \ll N$. However, it vanishes beyond $k_{\mathrm{max}}$ because 
the global maximum typically occurs in the shell $k_{\mathrm{max}}$. 
The average 
number ${\cal{R}}_{\mathrm{shell}}$ of records can be obtained by simply 
integrating $\tilde{P}_{\mathrm{shell}}(k,N)$ over $k$ and we find  
\be
{\cal{R}}_{\mathrm{shell}} \approx \frac{(\ell-\ln\ell-1)}{\ell-1} \; N \;\;.
\ee
Thus, ${\cal{R}}_{\mathrm{shell}}$ 
increases with $\ell$ and as $\ell \rightarrow \infty$, 
${\cal{R}}_{\mathrm{shell}} \rightarrow N$ because the population in each shell
becomes infinite. 
Since the $Y_{k}$'s are independent random variables with finite mean 
and variance, the number of records satisfies the central limit theorem
and becomes Gaussian for large $N$. Specifically, for large $N$, 
the variance of the number of records is
\be
\l{variance} {\cal{V}}_{\mathrm{shell}} = \sum_{k=0}^N 
\mbox{Prob}(Y_k = 1)[1 - \mbox{Prob}(Y_k = 1)]
\approx
\frac{N}{\ell - 1} 
\left( \frac{\ell + 1}{\ell - 1} \ln \ell - 2 \right) \;\;,
\ee
which is maximal for $\ell = 5$ and vanishes for large $\ell$.
The ratio ${\cal{V}}_{\mathrm{shell}}/{\cal{R}}_{\mathrm{shell}}$ is always 
considerably smaller
than unity, which implies that the distribution is much sharper than the
log-Poisson distribution obtained in the i.i.d. case.

\subsection{Inter-record spacings}
\l{recordspace}

For the discussion of the spacings between records, it is convenient
to label the records ``backwards'' in time, with $r_1$ denoting
the position of the last record (i.e., the global maximum), $r_2 < r_1$
the penultimate record, and so on.
In this way the pathologies associated with the fact that
the expected waiting time for the next record to occur is
infinite in the i.i.d. case \cite{Glick78} can be avoided.  
The probability $\tilde{P}(r_{j})$ that the $j$'th record
occurs at location $r_{j}$ 
can be found for the Nevzorov model in a manner analogous to 
(\ref{PYk12}). We obtain 
\be
\tilde{P}(r_{j})= 
\sum_{r_{1},..., r_{j-1}} \;  
\prod_{k=0,...,j-1} \frac{\alpha_{r_{k+1}}}
{\alpha_{0}+...+\alpha_{{r_{k}-1}}}  \;\;,\l{recloc}
\ee 
with $N \geq r_{1} >...> r_{j-1} > r_j$ and $r_{0}=N+1$. The factors on the 
right hand side of the above equation simply reflect the fact that 
the record at location $r_{k+1}$ remains the maximum 
until the next record occurs at $r_{k} > r_{k+1}$. 

For the i.i.d. model, using $\alpha_{k}=1$ for all $k$ in (\ref{recloc}),  
the average location $\av{r_{j}}=\sum r_{j} \tilde{P}(r_{j})$ can be 
calculated. One finds that the average inter-record distance 
$\tilde{\Delta}_{\mathrm{iid}}(j)=\av{r_{j}}-\av{r_{j+1}}$ 
between the $j$'th and $(j+1)$'th record behaves as \cite{Arnold98,Nevzorov01}
\be
\tilde{\Delta}_{\mathrm{iid}}(j) \approx \frac{N}{2} \left( \frac{1}{2} \right)^{j} 
\;\;,\;\;j=1,2,... \;\;. \l{interreciid}
\ee
A simple argument, useful for later discussion, can also be employed 
to obtain the above equation.
The record labeled by $j=1$, being the global maximum, is equally 
likely to occur 
anywhere between $1$ and $N$. Thus, on average, it is located at $N/2$. 
Similarly, the record labeled by $j=2$ is a global maximum in the 
range $[1,r_1)$ with 
uniformly distributed location, which gives the average location 
$\av{r_{2}}=(1/2) \av{r_1} = N/4$. Repeating this argument, we obtain the result in 
(\ref{interreciid}).

\begin{figure}
\begin{center}
\psfig{figure=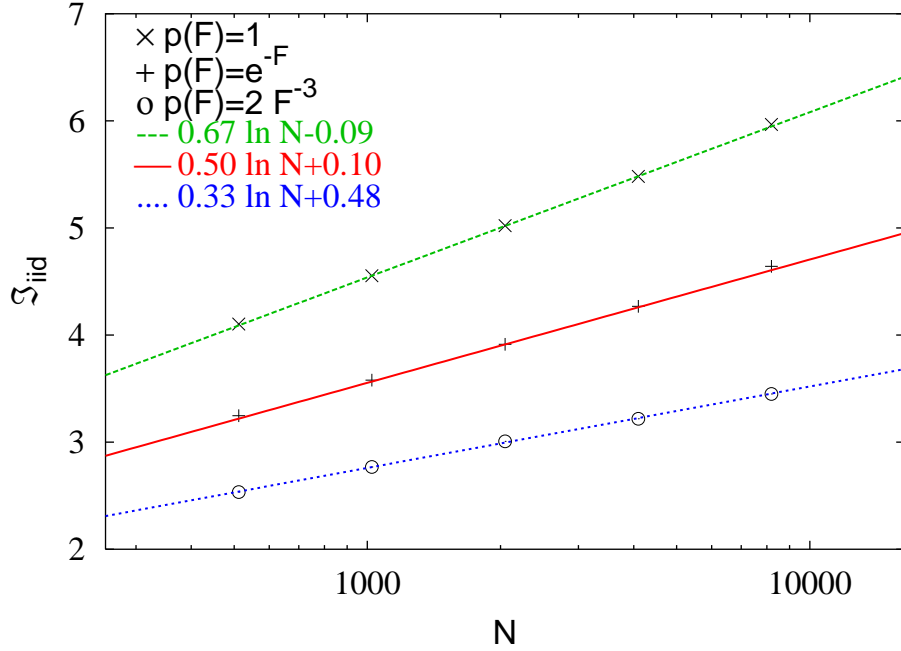,width=9cm,angle=270}
\caption{Mean number ${\cal{J}}_{\mathrm{iid}}$ of jumps for 
the i.i.d. model for 
various fitness distributions. The lines are the best fits to the functional 
form ${\cal{J}}_{\mathrm{iid}}= \beta \ln N + \mbox{constant}$.}
\label{iidjump}
\end{center}
\end{figure}

\begin{figure}
\begin{center}
\psfig{figure=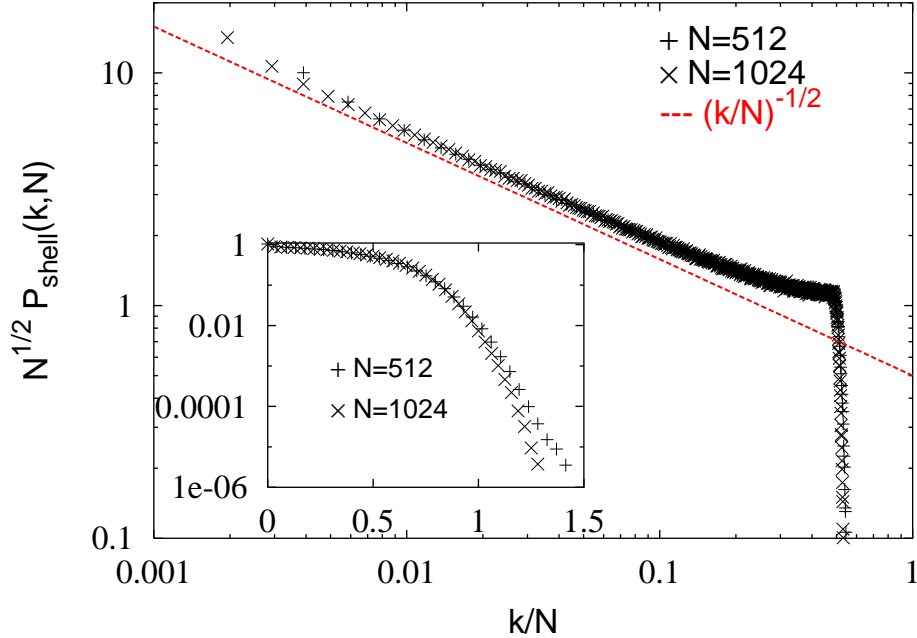,width=9cm,angle=270}
\caption{Data collapse for the scaled jump distribution 
$\sqrt{N} P_{\mathrm{shell}}(k,N)$ vs. $k/N$ for 
the shell model with $\ell=2$ and $p(F)=e^{-F}$. Inset: 
Scaled inter-jump spacing $\Delta_{\mathrm{shell}}(j)/\sqrt{N}$ vs. 
$j/\sqrt{N}$ to show that jumps are roughly equally spaced in the shell model.}
\label{sjump}
\end{center}
\end{figure}

\begin{figure}
\begin{center}
\psfig{figure=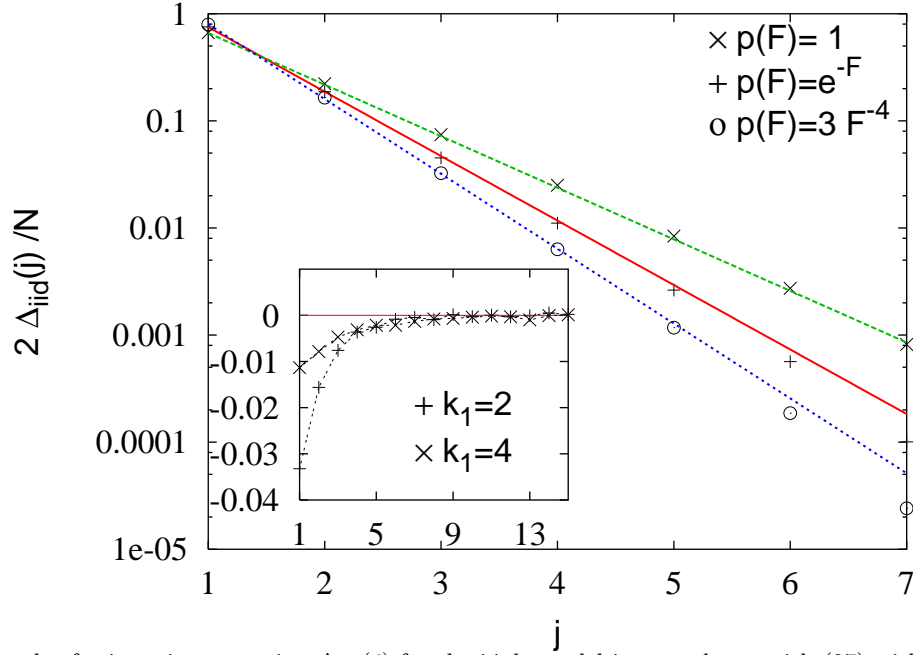,width=9cm,angle=270}
\caption{Semi-log plot for inter-jump spacing $\Delta_{\mathrm{iid}}(j)$ 
for the i.i.d. model in accordance with (\ref{interjumpiid}) with 
slope $b=\beta/2$. 
Inset: Correlation $C_{\mathrm{iid}}(k_{1},k_{2})$ 
vs. $k_{2}-k_{1}$ for two fixed values of $k_{1}$ 
to show correlations between jumps in the i.i.d. model.}
\label{rspace}
\end{center}
\end{figure}

For the shell model, since the most likely position of the 
global maximum is in the shell with the largest number $\alpha_k$ of 
sequences, we have $\av{r_{1}} = k_{\mathrm{max}}$. 
For sake of simplicity, we consider binary sequences ($\ell=2$)
in the following but the scaling behavior obtained below holds 
for $\ell > 2$ as well. We find 
that for $j \ge 2$, the average location $\av{r_{j}}$ of the 
$j$'th record is given by 
\be
\av{r_{j}}=\av{r_{1}} -\frac{1}{\pi} \sqrt{\frac{N}{2}} 
\left( \frac{2}{\sqrt{\pi}} \right)^{j-2} 
\int_{-\infty}^{\infty} dx_{j-1} \; 
\frac{e^{-x_{j-1}^{2}}}{\mbox{erfc}(x_{j-1})} \; \int_{x_{j-1}}^{\infty} 
dx_{j-2} \; \frac{e^{-x_{j-2}^{2}}}{\mbox{erfc}(x_{j-2})}  \; ... \;
\int_{x_{2}}^{\infty} 
dx_{1} \; \frac{e^{-2 x_{1}^{2}}}{\mbox{erfc}(x_{1})}  \;\;,\;\;j \ge 2 \;\;,
\l{rj}
\ee
where we have used Eqs.(\ref{cnr})-(\ref{avgcnr}) and 
performed a Gaussian integral. The average location $\av{r_{2}}$ 
of the second record is given by 
$\av{r_{2}}=\av{r_{1}}- 2.0064 \sqrt{N}/ \pi \sqrt{2}$. Thus, the second 
record (and in fact, $j$'th record for $j$ of order unity) can be found 
within ${\cal{O}}(\sqrt{N})$ distance of the global maximum since 
$\alpha_{k}$ has width $\sim \sqrt{N}$ about $k_{\mathrm{max}}$. 
For $j > 2$, 
after repeated integration by parts, (\ref{rj}) can be rewritten as 
\be
\av{r_{j}}=\av{r_{1}}-\frac{1}{\pi} \sqrt{\frac{N}{2}} \sum_{m=1}^{j-2} 
\frac{(\ln 2)^{j-2-m}}{(j-2-m)!} \;\; G(m) \;\;,\;\;j > 2 \;\;,\l{sspacing}
\ee
where 
\be
G(m)= (-1)^{m} \int_{-\infty}^{\infty} dx \; 
\frac{e^{-2 x^{2}}}{\mbox{erfc}(x)} \; 
\frac{\left( \ln {\mbox{erfc(x)}} \right)^{m}- 
\left( \ln {2} \right)^{m}}{m!} \;\;.
\ee
As outlined in Appendix {\ref{appendix1}, the integral $G(m)$ can be 
estimated by the saddle point method and we find 
$G(m) \approx \sqrt{ \pi m/2}$ for large $m$. Since the leading contribution 
to the sum in (\ref{sspacing}) comes from the $m=j-2$ term, we have 
\be
\tilde{\Delta}_{{\mathrm{shell}}}(j) \approx  
\sqrt {\frac{N}{4 \pi j}} \;\;,\;\; j \gg 1 \;\;. \l{interrecshell}
\ee
Thus, while the inter-record spacing decays exponentially with $j$ for 
the i.i.d. model, it falls as a power law for the shell model. 
The spacing between the first few records [$j = {\cal{O}}(1)$]
is of order $\sqrt{N}$, while for the bulk of the records with $j = 
{\cal{O}}(N)$ the spacing is of order unity; this is consistent with
the vanishing of the record occurrence probability (\ref{Ptildeshell})
near $k=k_{\mathrm{max}}$.

\section{Jump statistics}
\l{jump}

As we described in Sect.\ref{model}, in our model, evolutionary jumps
are a subset of records and if a jump occurs at 
$k^{\prime}$, the next jump is said to occur 
at $k > k^{\prime}$ if (i) $F(k)$ is a record (ii) the overtaking time 
$T(k, k^{\prime})= \mbox{min}_{j \geq k^{\prime}}  
\{ T(j, k^{\prime}) \}$. By convention, the first jump and the first record
occurs at $k=0$. Due to the second condition, some of the records can get 
bypassed and fail to appear in the set of jumps. 
In this section, we find the mean number of jumps and the 
inter-jump spacing for both the i.i.d and the shell model. 
In contrast to the properties of records, the statistics of jumps
depends explicitly on the fitness distribution $p(F)$ and we will consider 
distributions for which the tail behavior corresponds to the 
three universality classes of 
standard extreme value theory \cite{Sornette00} and which also 
appear in a very similar form in the theory of records 
\cite{Nevzorov87,Arnold98,Nevzorov01}. 

\subsection{Mean number of jumps}
\l{meanjump}

We begin by discussing numerical results for the average number of 
jumps in the i.i.d. model. 
It was found in \cite{Krug03} that the average number ${\cal{J}}_{\mathrm{iid}}$ 
of jumps grows as $\beta \ln N$ where the 
prefactor $\beta < 1$, and was conjectured to be 
\be
\beta \approx \cases { 1/2   & {$\;,\;\;p(F) \sim e^{-F}$} \cr
(\delta-1)/(2 \delta -1) & {$\;,\;\;p(F) \sim F^{-1-\delta}, \;\; \delta \geq 1$}  \cr
(2 + \nu)/(3 + 2 \nu)  & {$\;,\;\;p(F) \sim (F_{\mathrm{max}}-F)^{\nu}, \;\; \nu > -1$} 
} \;\;. \l{beta1d}
\ee
Figure~\ref{iidjump} shows ${\cal{J}}_{\mathrm{iid}}$ increasing linearly with 
$\ln N$ and slope $\beta$ for 
some distributions in accordance with (\ref{beta1d}). 
Thus, in the i.i.d. model
the jumps can be viewed as ``diluted'' records, in the sense that the mean 
number of 
records and jumps differ only up to a prefactor and  
the probability $P_{\mathrm{iid}}(k,N)$ that a jump occurs at $k$ 
is given by $\beta /k$. In this picture, the probability for a given
record to be bypassed is simply $1-\beta$. However, bypassing is not
completely random, as the variance of the number of jumps is found
consistently to be smaller than the mean. This implies a certain amount
of ``anti-bunching'' among the jumps, which can also be detected
by the direct measurement of correlation 
$C_{\mathrm{iid}}(k_{1},k_{2})=P_{\mathrm{iid}}(k_{1},k_{2})-
P_{\mathrm{iid}}(k_{1}) P_{\mathrm{iid}}(k_{2})$  
where $P_{\mathrm{iid}}(k_{1},k_{2})$ is the joint distribution of 
having a jump at $k_{1}$ and $k_{2}$, as shown in the inset 
of Fig.~\ref{rspace}. 
Some further discussion of the conjecture (\ref{beta1d}) will be provided
in Sect.\ref{beta-z}.

Estimates for the mean number of jumps in the shell model were also
given in \cite{Krug03}, but the range of sequence lengths was too limited
to allow for a definite statement. Here we present the results of our 
simulations for large values of $N$ obtained using the approximation 
described below. Since the shell fitness $F(k)$ is the 
maximum of $\alpha_k$ i.i.d. random variables drawn from the distribution
$p(F)$, it can be obtained from a uniform random variable 
$u$ using the relation   
\be
\l{uniform}
\int_{F_{\mathrm{min}}}^{F(k)} p(x) dx =u^{1/\alpha_{k}}.
\ee 
However, the binomial coefficient ${N \choose k}$ increases  
exponentially with $N$ for 
large $k$ and the distribution of $u^{1/\alpha_{k}}$ approaches a 
delta-function centred at unity for $k \gg 1$ making it difficult to determine 
the distribution $p_{k}(F)$ of $F(k)$ accurately when $N$ is large. 
For the exponential fitness distribution, the relation (\ref{uniform})
becomes 
\be
\l{Fexp}
F(k)=-\ln (1- e^{\ln u/\alpha_{k}}) \simeq - \ln [\ln(1/u)]+\ln(\alpha_{k}).
\ee
Since the last expression only involves the logarithms of binomial 
coefficients, $F(k)$ can be easily generated up to large values of $N$.  
While a similar approximation can be employed for other 
distributions with unbounded tails, we have not been able 
to obtain reliable results for bounded distributions.  

In Fig.~\ref{sjump}, we show simulation results for the
probability $P_{\mathrm{shell}}(k,N)$ that a jump occurs in 
the $k$'th shell, for a binary alphabet  
($\ell=2$) and two different values of $N$. The data collapses 
onto the scaling form 
\be
P_{\mathrm{shell}}(k,N) \approx  N^{-1/2} f(k/N) \;\;\;,\;\; 
x < \frac{\ell-1}{\ell} = \frac{k_{\mathrm{max}}}{N} \;\;,
\l{scaling_jumps}
\ee
where the scaling function behaves as $f(x) \sim x^{-1/2}$ for small $x$.
This has the interesting consequence that 
$P_{\mathrm{shell}} \sim 1/\sqrt{k}$ is independent
of $N$, and hence the number of jumps grows as $\sqrt{k}$ 
for $k \ll k_{\mathrm{max}}$.
The scaling function appears to be independent of the alphabet size $\ell$,
which therefore only changes the cutoff at $k = k_{\mathrm{max}}$ \cite{Krug04}.
Thus, the average number ${\cal{J}}_{\mathrm{shell}}$ of jumps obtained by 
integrating $P_{\mathrm{shell}}(k,N)$ over $k$ grows with $N$ as 
\be
{\cal{J}}_{\mathrm{shell}} \sim \sqrt{\frac{(\ell -1) N}{\ell}} \;\;\;.
\ee
Our simulations indicate that the above dependence of 
${\cal{J}}_{\mathrm{shell}}$ on $N$ is also true for 
normal-distributed fitness and we expect it to hold 
for all distributions decaying more rapidly than 
any power law \cite{Krug04}. Further, the distribution of the number of jumps 
is a Gaussian (as in the case of records) with both mean and variance 
scaling as $\sqrt{N}$.  

For the power law case, we find that ${\cal{J}}_{\mathrm{shell}} \to 1$ for 
large $N$. 
As we shall see in Section~\ref{approach}, the globally fittest sequence 
takes over the leadership in a time of order unity in this case, 
which explains the above behavior of ${\cal{J}}_{\mathrm{shell}}$. 
In summary, unlike the i.i.d. model, most of the records are 
bypassed in the shell model both for exponential and power law distributions.

\subsection{Inter-jump spacings}
\l{inter-jump}

We now turn to a discussion of the inter-jump spacings 
$\Delta(j)$ 
defined in analogy to the inter-record spacings.
We denote by $s_j$ the position of the $j$'th jump, with
$j=1$ referring to the last jump, which is also the last
record (the global maximum), and define the inter-jump spacing
as $\Delta(j) = \av{s_{j}} - \av{s_{j+1}}$ where $\av{s_{j}}$ is the average 
location of the $j$'th jump. 
For the i.i.d. model, an approximate calculation of  
$\Delta_{\mathrm{iid}}(j)$ can be carried out by assuming jumps to be 
randomly diluted records.
The position $s_j$ of the $j$'th jump equals the position
$r_k$ of the $k$'th record, where $k \geq j$ because of the possibility of
bypassing. If the record $k+1$ is not bypassed, then $s_{j+1} = r_{k+1}$
and $s_{j+1} = (1/2) s_j$ on average due to the argument 
given after (\ref{interreciid}) for the inter-record spacings.
In the diluted record picture,
this is true with probability $\beta$. With probability $\beta (1 - \beta)$,
record $k+1$ is bypassed and $s_{j+1} = r_{k+2}$ so that 
$s_{j+1} = (1/4) s_j$ on the average. Similarly, with
probability $\beta (1 - \beta)^l$, $l$ records are bypassed
and the ratio $s_{j+1}/s_j = (1/2)^{l+1}$ on average. Summing over
all possibilities, we find that $\av{s_{j+1}} = b \av{s_j}$ with
\be
\l{bgeom}
b = \sum_{l = 0}^\infty 2^{-(l+1)} \beta (1 - \beta)^l = \frac{\beta}{1 + \beta}.
\ee
Taking into account that the sum over all inter-jump spacings adds up to 
$\av{s_1} = \av{r_1} = N/2$, we obtain
\be
\Delta_{\mathrm{iid}}(j) \approx \frac{N}{2} (1- b) b^{j-1}
\;\;,\;\;j=1,2,... \;\;.\l{interjumpiid}
\ee
The numerical data shown in Fig.~\ref{rspace} supports the general
form of this expression, but with the coefficient 
$b \approx \beta/2$ rather than $\beta/(1+\beta)$. This is another
indication of correlations that are not accounted for in the random
dilution picture.

For the shell model, as shown in Fig.~\ref{sjump}, the data 
for $\Delta_{\mathrm{shell}}(j)$ for various $N$ collapses onto a 
monotonically decreasing curve if we assume $\Delta_{\mathrm{shell}}(j)$ 
to be of the scaling form 
\be
\Delta_{\mathrm{shell}}(j) \approx 
\sqrt{N} \; h(j/\sqrt{N}) \;\;. \l{interjumpshell}
\ee
The scaling with $\sqrt{N}$ follows naturally from the scaling form 
(\ref{scaling_jumps}) for the jump occurrence probability. The latter
implies that the density of jumps on the $k$-axis is of order $1/\sqrt{N}$
for finite $k/N$, hence the spacing is of order $\sqrt{N}$, and the
argument of the scaling function is $j/\sqrt{N}$ because the total number
of jumps is also of order $\sqrt{N}$. However, the tail of the scaling 
function $h$ does not obey the scaling form (\ref{interjumpshell}) and 
approaches zero with increasing $N$. Thus, while for finite $k/N$, the jumps 
are roughly equally spaced with spacing $\sqrt{N}$, the spacing is of 
${\cal{O}} (N^{s})$ with $s < 1/2$ for $k/N \rightarrow 0$.


\section{Approach to the global fitness maximum}
\l{approach}

So far, we have discussed the statistics on the $k^{\ast}$ axis of the 
inset of Fig.~\ref{lines} and now we focus on the temporal statistics.   
As explained in Section~\ref{model}, a fit sequence $k^{\ast}$ leads till 
it is 
overtaken by an even fitter one and eventually the globally fittest sequence 
emerges as the leader at typical time $T^{\ast}$. 
In this section, we find this typical time $T^{\ast}$ required 
for the population to reach the global maximum of the fitness landscape 
and the distribution of the evolution times to jump from one local maximum to 
another. 

\subsection{Dynamic scaling}

The location $k^{\ast}(t)$ of the most populated sequence at time $t$ for 
which the logarithmic population is 
$E({k^{\ast}},t)=\mbox{max}_{k} \{ E(k,t) \}$ increases with time till 
the global maximum is reached. In Appendix~\ref{appendix2}, the 
distribution $P_t(k^{\ast})$ is explicitly calculated for the 
i.i.d. model in the limit of infinite genome size ($N \to \infty$). 
This distribution is usually of the scaling form 
\be
\l{Ptscale}
P_t(k^\ast) = t^{-1/z} \Phi(k^\ast/t^{1/z}) \;, \l{Ptkast}
\ee
where the scaling function $\Phi$ depends on the underlying fitness 
distribution
and the dynamic exponent $z$ is given by 
\be
z= \cases { 1   & {$\;,\;\;p(F) \sim e^{-F}$} \cr
  (2 + \nu)/(1 + \nu)  & {$\;,\;\;p(F) \sim (F_{\mathrm{max}}-F)^{\nu}, \;\; \nu > -1$} \cr
     (\delta-1)/\delta & {$\;,\;\;p(F) \sim F^{-1-\delta}, \;\; \delta \geq 1$}
} \;\;, \l{1dz}
\ee
for the three classes of fitness distributions introduced in Sect.\ref{jump}. 
The corresponding behavior $\overline{k^{\ast}(t)} \sim t^{1/z}$ 
of the mean location has been derived previously using Flory-type arguments 
\cite{Zhang86,Krug93} and can also be seen using (\ref{Ptkast}). As we have 
already discussed, the global maximum is reached
when $\overline{k^{\ast}(t)} \approx N/2$, which defines a total evolution 
time $T^\ast \sim N^z$. One thus expects a scaling form 
\be
\overline{k^{\ast} (t,N)} \approx t^{1/z} \varphi (t/T^{\ast}) \;\;,
\ee
where the scaling function $\varphi(x)$ is a constant for $x \ll 1$ and 
decays as $x^{-1/z}$ for $x \gg 1$ \cite{remark}.

An alternative approach \cite{Krug02,Krug03} to estimating 
the typical time $T^{\ast}$ required by the population 
to reach the global maximum starts from the observation that
$T^\ast$ is of the order of the time $T_1$ at which 
the globally fittest sequence at typical location $r_{1} = s_1$ and fitness 
$F(r_1)$ 
overtakes the penultimate leader with respective quantities $s_2$ and 
$F(s_2)$ (refer Sect.\ref{inter-jump}). From (\ref{Tkkprime}), we have
\be
T_1 =  \frac{s_1 - s_2}{F(s_1)-F(s_2)} \;\;,
\l{Tast}
\ee
where the inter-jump spacing in the numerator is given by 
(\ref{interjumpiid}). 
The estimation of the denominator involves a subtlety -- in previous 
works \cite{Krug02,Krug03}, it was assumed that $F(s_1) - F(s_2)$ is of the
order of the \textit{fitness gap} $\epsilon$, which was defined as 
the difference between the values of the global maximum 
and the second largest fitness of the fitness landscape. 
Because the second largest fitness does not necessarily appear in the 
record sequence,
$\epsilon$ is only a lower bound on $F(r_1) - F(r_2)$, which in turn is clearly
a lower bound on the fitness difference of interest,
$\epsilon \leq F(r_1) - F(r_2) \leq F(s_1) - F(s_2)$. However, our 
explicit calculations show that $F(r_1) - F(r_2)$ is of the 
same order as $\epsilon$; moreover, at least for the i.i.d. model, we know 
that at most a few
records are bypassed between $s_1$ and $s_2$, and hence the assumption
that $F(s_1) - F(s_2) \sim \epsilon$ seems justified.

The calculation of the distribution of fitness gap $\epsilon$ is a 
standard exercise in
extreme value statistics \cite{David70}. 
In a system with a total number $S$ of sequences, the typical value of the 
fitness gap increases as $S^{1/\delta}$ for the unbounded power 
law distribution, decreases as $S^{-1/(1+\nu)}$ for the bounded distribution,
and is of order unity for the exponential distribution. 
For the i.i.d. model, using 
$s_{1} - s_{2} \sim N$ [see (\ref{interjumpiid})] and $S=N$ in 
(\ref{Tast}), we recover $T_1 \sim T^{\ast} \sim N^{z}$ with $z$ given in 
(\ref{1dz}). The other fitness distributions can be treated in a similar 
manner. 

For the shell model, $S=\ell^{N}$ and due to (\ref{interjumpshell}), the 
numerator is of order $\sqrt{N}$ so that $T^{\ast} \sim \sqrt{N}$ for 
the case of exponentially distributed fitness. Presumably, the time $T_{j}$ 
at which the $j$'th jump occurs is also of order $\sqrt{N}$ for 
$j \sim {\cal{O}}(1)$. Since the total number of jumps is of the same order, 
it follows that initially there are many, quick jumps followed by few 
jumps that take ${\cal{O}} (\sqrt{N})$ time. 
This result agrees qualitatively with that seen 
in experiments (discussed later) concerning the pace of evolution which is 
initially rapid and later slows down considerably. 
For the bounded distributions, $T^{\ast}$ increases exponentially with $N$ 
whereas it \textit{decreases} exponentially for the 
fat-tailed power law distributions. The latter result implies 
that, for large $N$, the global maximum
takes over in a single time step, which explains why the mean number of 
jumps tends to unity  for the power 
law distributions (see Sect.\ref{meanjump}).

\subsection{Universal tails of the evolution time distribution}

In the last subsection, we found the typical time $T^{\ast}$ to 
reach the global maximum and now we 
consider the distribution $P(T_{1},N)$ of the time $T_{1}$ 
at which the final jump occurs. For the i.i.d. model, since the 
typical $T_{1}$ also grows 
as $N^{z}$, we may expect the normalised distribution 
$P(T_{1},N)$ to be of the scaling form 
$P(T_{1},N) \approx N^{-z} g_{1}(T_{1}/N^{z})$.
In general, for $j \sim {\cal{O}}(1)$, the distribution $P(T_{j},N)$ of the time 
$T_{j}$ at which the $j$'th jump occurs is of the scaling form 
\be
\l{Tjscale}
P(T_{j},N) \approx N^{-z} g_{j}(T_{j}/N^{z}) \;\;.
\ee
Although the dynamic exponent 
$z$ depends on the underlying fitness distribution, we shall now show 
that the tail of the 
distribution $P(T_{j},N)$ is universal. The events contributing to 
large $T_{1}$ are the ones for which $F(s_1) - F(s_2)$ is small; 
in these cases we expect the general bound $F(s_1) - F(s_2) \geq F(r_1) - F(r_2)$
to be saturated, i.e. the second record is not bypassed and $s_2 = r_2$. 
Thus, using (\ref{Tast}), we obtain
\be
\l{PT1}
P(T_{1},N) \approx \left| \frac{d \epsilon_1}{d T_1} \right|
\mbox{Prob}(\epsilon_1 = \Delta_{\mathrm{iid}}(1)/T_1) \approx
\frac{\Delta_{\mathrm{iid}}(1)}{T_{1}^2}  \; 
\mbox{Prob}(\epsilon_{1}=0) \;\;, 
\ee
for large $T_1$,
where $\epsilon_{1}=F(r_{1})-F(r_{2})$. 
The probability distribution of $\epsilon_1$ can be obtained along
the lines of the derivation of the distribution of the fitness gap
$\epsilon$ in \cite{Krug02}, and it is found that 
the probability for a near-vanishing difference between two successive record 
values is nonzero for any $p(F)$.
We conclude that $P(T_{1},N)$ has a 
power law tail with 
exponent $-2$ for any underlying fitness distribution \cite{Krug03}.
This is an example of the generation of a power law through a change
of variables (from $\epsilon_1$ to $T_1 \sim 1/\epsilon_1$) as described
in \cite{Sornette00}. 

Similarly, the events contributing to the tail of $P(T_{2},N)$ 
are the ones in which the record at $r_{2}$ is the penultimate leader and 
the record at $r_{3}$ is the leader previous to it. Thus, we demand that 
none of these two records should be bypassed; in particular,
$r_2$ should not be bypassed, which requires 
that $T(r_{1}, r_{2}) > T(r_{2},r_{3})$.  
This condition can be written as 
$\epsilon_{1} < C \epsilon_{2}$, where
$\epsilon_2 = F(r_{2})-F(r_{3})$ and $C$ is the average value
of $(r_1-r_2)/(r_2-r_3)$, a number of order unity. Thus we obtain 
\be
P(T_{2},N) \approx  \frac{\Delta_{\mathrm{iid}}(2)}{T_{2}^2}  
\int_{0}^{C \Delta_{\mathrm{iid}}(2)/T_2} d \epsilon_{1} \; 
\mbox{Prob}(\epsilon_{1}, \epsilon_{2},N)  \;
\approx \frac{C \Delta_{\mathrm{iid}}(2)^2}{T_2^3}
\mbox{Prob}(\epsilon_{1}=0, \epsilon_{2}=0,N).  
\ee
Since Prob$(\epsilon_{1}=0, \epsilon_{2}=0,N)$ can be shown to be 
nonzero, we conclude  
that $P(T_{2},N) \sim T_2^{-3}$ for large $T_2$. 
Extending the above arguments in a similar fashion to the next evolution 
times, we find that the scaling functions in (\ref{Tjscale}) behave as
$g_{j}(x) \sim x^{-1-j}$ for $x \gg 1$. 
Interestingly, this implies that the expected time $\av{T_j}$ is finite
for $j \geq 2$ and infinite for $j = 1$.
In Fig.~\ref{evol}, the prediction 
$P(T_{j},N) \sim T^{-1-j}$ for $j=1,2,3$ is compared with data 
obtained using Monte Carlo simulations for the i.i.d. model. 
The behavior of the universal tails of $P(T_{j})$ discussed above 
is true for the shell model as well.

\section{Bypassing probability and dynamic exponent: a conjecture}
\l{beta-z}

As we have seen in the previous sections, the jump statistics are not 
analytically tractable due to the constraint of minimal overtaking time. 
For the i.i.d. model, the jumps differ from the records only up to a prefactor 
$\beta$ conjectured to be given by  (\ref{beta1d}). 
Comparing the expressions in Eqs.(\ref{beta1d}) and (\ref{1dz}), we observe 
that 
the bypassing probability $1- \beta$ appears to be related to the 
dynamic exponent $z$ by the following universal relation 
\be
\beta = (1-\beta) \; z \;\;. 
\l{relation}
\ee
A derivation of this relation (which eludes us so far) would constitute a 
proof of the conjecture (\ref{beta1d}). 

Interestingly, the relation (\ref{relation}) can be interpreted in terms of a 
kind
of \textit{duality} between the $k$- and the $t$-axis of the inset of 
Fig.~\ref{lines}. So far we have identified each jump with the position on the
$E$-axis where the line that takes over the leadership when the jump
occurs originates; but we may just as well identify the jump with the 
corresponding crossing time $T(k,k^{\prime})$ at which the leadership 
shifts from $k'$ to $k$. 
Clearly, there is a one-to-one correspondence between the jumps defined on 
two axes and the average number ${\cal{J}}_{\mathrm{iid}}^{\prime}$ of jumps 
on the time axis is equal to ${\cal{J}}_{\mathrm{iid}}$ discussed in earlier 
sections. 
Thus, ${\cal{J}}_{\mathrm{iid}}^{\prime}={\cal{J}}_{\mathrm{iid}} 
\approx \beta \ln {N}$ and 
since the 
typical time $T^{\ast}$ to reach the global maximum scales as $N^{z}$, we have 
${\cal{J}}_{\mathrm{iid}}^{\prime} \approx \beta^{\prime} 
\ln {T^{\ast}}$ with $\beta^{\prime}= 1- \beta$ due to the conjecture 
(\ref{relation}).  
This leads us to expect that the probability 
$P_{\mathrm{iid}}(t,N)$ that a jump occurs at time $t$ decays as 
$\beta^{\prime}/t$ for $t \ll N^{z}$ and as 
$1/t^{2}$ for $t \gg N^{z}$; the latter behavior is the universal 
$t^{-2}$ tail explained in Section~{\ref{approach}}. 
We conclude that 
the sum of the jump probabilities along the $k$- and the $t$-axes should
sum up to a universal function, 
{\it i.e.}
\be
P_{\mathrm{iid}}(t=X,N)+P_{\mathrm{iid}}(k=X,N) = 1/X \;\;,
\ee
for any choice of fitness distribution. 
The numerical evidence supporting this claim is shown in Fig.~\ref{sum}.
Furthermore, in analogy to the jump spacing $\Delta_{\mathrm{iid}}(j)$ along 
the $k$-axis, 
one can also consider the quantity 
$\Delta_{\mathrm{iid}}^{\prime}(j) = \langle T_j \rangle - 
\langle T_{j+1} \rangle$ 
which is the spacing 
between the successive jumps on the time-axis. Replacing $N$ by $T^{\ast}$ 
and $b = \beta/2$ by $\beta'/2 = (1-\beta)/2$ in (\ref{interjumpiid}), we 
expect
\be
\Delta_{\mathrm{iid}}^{\prime}(j) \sim N^{z} 
\left( \frac{1-\beta}{2} \right)^{j-1} \;\;,\;\;j=1,2,...  \;\;. \l{cnjctre}
\ee
Numerical results consistent with this expression 
are shown in the inset of Fig.~\ref{sum} for some distributions. 
The deviations seen in the data for the first jump ($j=1$) reflect the 
fact that,
because of the $1/T_1^2$-tail derived in Eq.(\ref{PT1}), the average of $T_1$
is not defined and hence grows with the number of disorder realizations. 

\section{A Model Based on Record Times}
\l{mft}

In this section, we introduce a further simplification of the i.i.d. model. 
As we have discussed already, a sequence $k$ may occur in the set of 
jumps provided $F(k)$ is a record. Thus, it is sufficient to consider 
only the subset of sequences whose fitness is a record.
Here it is convenient to label the records forward in time, so we
denote by $R_j$ the location of the $j$'th record with $j=1$ labeling
the first record ($R_1 = 1$ by convention), $R_2 > R_1$ the second record,
and so on.  Note that 
there are two sources of randomness in the problem -- one arising from the
record locations $R_j$ and the other due to record 
values $F(R_j)$. For exponentially distributed fitness, it is known that 
the differences between 
successive record values are independent and exponentially distributed 
random variables \cite{Tata69}. Thus, the fitness of two  
successive records differs by unity on average. 
These considerations allow us to 
eliminate the randomness associated with the record values by replacing
the i.i.d. model of (\ref{shell}}) with exponentially distributed fitness by 
a simpler model for which the population evolves as 
\be
\tilde E(j,t)=-R_{j}+ j t.
\label{mfmodel}
\ee
Like in the original i.i.d. model, we find numerically that the 
average number $\tilde {\cal{J}}$ of jumps grows logarithmically with $N$
with a prefactor $\tilde \beta \approx 0.63$ (see Fig.~\ref{mftjump}). 
This is distinctly different from the value $\beta \approx 1/2$ found for 
the i.i.d. model with exponential fitness distribution, indicating that the 
randomness in the record values is relevant. Somewhat surprisingly, in 
contrast to the conjectured values of $\beta$ for the full i.i.d. problem 
given in (\ref{beta1d}), $\tilde \beta$ does not 
seem to be a simple rational number.

Our primary motivation for introducing this simplified model is to gain further
insight into the mathematical structure of bypassing. 
For the model defined by (\ref{mfmodel}), 
the crossing time $\tilde T(j,j^{\prime})$ at which the line associated 
with the $j$'th record overtakes that associated with record $j'$ is given by
\be
\label{mftimes} 
\tilde T(j,j')= \frac{R_{j}-R_{j'}}{j-j'}.  
\ee
Then the probability $\tilde \beta_{2}$ that the second record is not
bypassed can be written as 
\bea
\tilde \beta_{2} &=& 
\mbox{Prob} \left[ R_{2}-1= \mbox{min} \left( R_{2}-1, \frac{R_{3}-1}{2}, 
\frac{R_{4}-1}{3},...\right) \right].  
\eea
The evaluation of the 
condition on the right hand using the joint 
probability distribution for the record 
times \cite{Nevzorov87,Arnold98,Nevzorov01}
\be
\label{jointRn}
\mbox{Prob}(R_2,R_3,...,R_n) = 
\frac{1}{(R_2 - 1)(R_3 - 1)(R_4 - 1)...(R_n - 1)R_n}
\ee
is clearly a difficult task. An upper bound on $\tilde \beta_2$ is obtained by 
requiring that the record at $R_2$ is not bypassed by the one at $R_3$, i.e.
that $\tilde T(2,1) < \tilde T(3,1)$. This gives  
\be
\label{b2bound}
\tilde \beta_2 \leq 
\mbox{Prob}(R_{3} > 2 R_{2} - 1)= \sum_{R_{2}=2}^{\infty} 
\frac{1}{R_{2}-1} \; 
\frac{1}{2 R_{2}-1} = 2 \; (1- \ln 2) \approx 0.613706 \;\;.
\ee
In our simulations on a large system, we find 
$\tilde \beta_{2} \approx 0.600786$, showing that bypassing of $R_2$
by the records beyond $R_3$ is rather unimportant. 

Consider next the behavior of the crossing times (\ref{mftimes}) when $j$ and
$j'$ are large. Williams \cite{Williams73} has shown that the sequence of 
record times
can be generated from the recursion relation 
\cite{Glick78,Nevzorov87,Arnold98,Nevzorov01}  
\be
\label{Williams}
R_{j+1} =  [ e^{X_j} R_{j}] + 1,
\ee
where the $X_j$ are independent, exponentially distributed random variables 
with mean one,
and $[ a ]$ is the integer part of $a$. For large $j$, the integer constraint 
can be ignored, and hence 
\be
\label{Tijlarge}
\tilde T(j,j') \approx \frac{R_{j'}}{j - j'} 
\left[ \exp \left(\sum_{i=j'+1}^j X_i \right) - 1
\right].
\ee
Recalling that 
the choice of the next non-bypassed record involves finding the minimum among
all crossing times $\tilde T(j,j')$ with $j > j'$, we see that that the current
location $R_{j'}$ cancels in the comparison between two such crossing times.
The problem thus acquires translation invariance in the 
record space, in the sense that the position of the next non-bypassed
record depends only on $j - j'$ and on the random
variables $X_i$ associated with the records between $j'$ and $j$. It is 
therefore plausible
that the bypassing probability tends to a constant for large $j$, and one is 
tempted
to describe the process by a Markov chain on the
set of records with the transition probability 
\be
\label{Pjk}
P_{j',j} = \mbox{Prob}[ \tilde T(j,j') = \min_{n > j'} \tilde T(n,j')].
\ee     
This is the conditional probability that the next jump occurs at $j$, given 
that
the preceding jump was at $j'$, averaged over all realizations of record times.
Using the representation (\ref{Tijlarge}), the $P_{j,j'}$ are manifestly 
translationally
invariant for large $j,j'$, depending only on $j-j'$.
Even in the asymptotic limit in which the expression (\ref{Tijlarge}) can be 
used,
the evaluation of (\ref{Pjk}) is cumbersome, but an analytic 
upper bound on $ P_{j,j+1}$ can be obtained along the lines of 
(\ref{b2bound}).  
In the limit of large $j$, we can write
\be
\label{Pjj+1bound}
P_{j,j+1} \leq \mbox{Prob}[\tilde T(j+1,j) < \tilde T(j+2,j)] 
= \mbox{Prob}[e^{X_{j+1}} + e^{-X_j} > 2] = \ln 2,
\ee
where $X_j$ and $X_{j+1}$ are the independent exponential random variables 
used in the 
representation (\ref{Williams}). The numerical evalution of (\ref{Pjk}) yields
$P_{j,j+1} \approx 0.669$, $P_{j,j+2} \approx 0.225$ and 
$P_{j,j+3} \approx 0.075$,
indicating a roughly exponential decay of the transition probability. 
From the $P_{j',j}$, the mean density of jumps (non-bypassed records) can be 
computed according to 
\be
\label{betamf2}
\tilde \beta_{\mathrm{Markov}} = 
\left( \sum_{n=1}^{\infty} n P_{j,j+n} \right)^{-1} \approx 0.676,
\ee
which is significantly larger than the direct numerical estimate 
$\tilde \beta \approx 0.63$
(see Fig.~\ref{mftjump}). This shows that the transition probability 
(\ref{Pjk}) is 
\textit{not} an exact representation of the process. The reason is that 
(\ref{Pjk})
is an \textit{annealed} average, whereas in the full problem the record times
(or, equivalently, the exponential random variables in (\ref{Williams}))
must be 
treated as \textit{quenched}: Minimizing the crossing times $\tilde T(j,j')$
for a given $j'$ involves, in principle, \textit{all} $j > j'$, and 
the \textit{same} set of 
random variables is used every time this minimization is repeated for 
different $j'$.

We thus have to conclude that the range of attainable analytic results, even 
for the 
simplified problem (\ref{mfmodel}), is very limited. An extension of the bound
(\ref{b2bound}) to the full i.i.d. model should be feasible using the 
representation
of the joint distribution of record times and record values through a Markov  
chain \cite{Arnold98,Nevzorov01}; however, as such a bound is unlikely to 
provide much insight
into the conjectured relation (\ref{relation}), we have not pursued this 
approach.

\section{Conclusions}
\l{conclude}

In this article, we characterised the evolutionary trajectories 
traced out by a quasispecies population in an uncorrelated rugged 
fitness landscape. These trajectories approach the global fitness maximum 
through a sequence of \textit {jumps} 
which mark a change in the identity of the most populated genotype. 
The statistics of these evolutionary jumps 
was studied mainly numerically. However, useful insights were provided by 
a study of \textit{record} statistics which could be handled analytically. 
It was found that the jump statistics are qualitatively similar to records, 
but there are quantitative differences because, as shown in 
Fig.~\ref{lines},   
a record breaking genotype can be \textit{bypassed} 
by a superior one before it can acquire dominance 
in the population (i.e. qualify to be a jump). 

The statistics of records and jumps depends strongly on the 
geometry of the space of genotypes. The natural setting for genotype 
evolution is the Hamming space of sequences of fixed length $N$. However, 
computational effort could be greatly reduced by lumping together 
the sequences within a shell of constant Hamming distance with respect to 
the initial population \cite{Krug03}. Complementary to this shell model, we 
also considered
a model of i.i.d. shell fitnesses, which corresponds effectively to a 
one-dimensional sequence space. While for the i.i.d. model, the average 
number of jumps differs from the number of records only through the 
prefactor $\beta$ of the logarithm of $N$, for the shell model
the ratio of the two numbers 
${\cal{J}}_{\mathrm{shell}}/{\cal{R}}_{\mathrm{shell}} \rightarrow 0$ as 
$N \rightarrow \infty$. 
For fat-tailed fitness distributions the evolutionary
trajectories in the shell model may even degenerate, in the sense that the 
global fitness maximum is reached in a single step. For distributions
decaying faster than a power law, like the exponential and normal 
distributions,
we find numerically that ${\cal{J}}_{\mathrm{shell}} \sim \sqrt{N}$; 
an analytic understanding of this result would be very desirable. 

A universal feature of the
evolutionary trajectories, which is independent of the geometry of genotype
space as well as of the fitness distribution, is the hierarchy of power 
law tails
for the distributions of the times at which the jumps occur. In particular, 
the $T^{-2}$-tail for the total evolution time implies that the average
of $T$ is infinite. The dependence of the 
\textit{typical} evolution time on the size of the sequence space
can be characterized by a dynamic exponent $z$, which was obtained 
exactly in terms of the extremal properties of the fitness distribution.
On a mathematical level, perhaps the most intriguing result of this work
is the conjectured relation (\ref{relation}) between  $\beta$ and $z$ for the 
i.i.d. model. It would be extremely interesting to understand how such a 
simple relation arises from the properties of the matrix of crossing 
times (\ref{Tkkprime}); however,
in view of the difficulties encountered even in the analysis of the simplified 
problem (\ref{mfmodel}), we do not see at present how further progress in this
direction can be achieved.

In this article, we worked in the limit of infinite population and strong 
selection. However, we expect our results to 
hold, at least qualitatively, for finite selective temperature as well. The
jumps appear instantaneous in the strong selection limit;
at finite selective temperature 
they take a finite amount of time in which the peak
associated with the new leader catches up with the currently dominating
peak and the population distribution briefly becomes bimodal 
\cite{Rujan88,Ebeling84,Krug93}.   
Although the quasispecies theory, which works in the infinite population 
limit, is inadequate 
to address the fluctuation effects that become important
when small mutant populations cross a fitness valley \cite{vanNimwegen00},   
for the related problem of episodic behavior in evolutionary
computation \cite{vanNimwegen97}, some of the finite
population behavior has been understood on the basis of the infinite population
limit. Thus we expect our investigation to give some insight into the
behavior of finite populations in rugged fitness landscapes 
\cite{Campos02a,Campos02b} as well. 

We close with some remarks on the applicability of the present work to the 
evolution of
biological populations. A realization of asexual mutation-selection dynamics 
in a 
static fitness landscape that is believed to be quite rugged is provided by
the long-term experiments on populations of \textit{Escherichia coli} carried
out by Lenski and collaborators \cite{Lenski94,Elena96,Elena03}. These 
experiments
show evidence for punctuated behavior both in the fitness and in the 
morphological
features (such as cell size) of the evolving populations, which is 
attributed to the
emergence and fixation of beneficial mutations. Fixation implies that a 
mutation which
is initially present only in a single individual is inherited by a growing 
fraction
of the population and eventually acquires dominance. 

At least on a qualitative level,
this process corresponds in our model to that in which a subpopulation 
residing in a distant shell
of sequence space takes over the leadership of the population, 
as described by (\ref{lineq}) and illustrated in Fig.~\ref{lines}. 
The bypassing of one subpopulation by another is analogous to the phenomenon of
\textit{clonal interference}, in which one beneficial mutation is superseded by
another one before reaching fixation \cite{Gerrish98}. 
The key difference between our model and the behavior of real asexual 
evolution is that in the latter case beneficial mutants arise through the 
stochastic search of a finite
population in an immensely large sequence space, while in the quasispecies 
model \textit{all possible mutants are present} 
(in extremely small numbers) \textit{after the first time
step}. It is, therefore, mandatory to include the 
stochastic finite population dynamics
into the model. Nevertheless, some features of the competition between 
the mutant populations (however they may have arisen) could 
well survive in a more complete, realistic treatment.   

\section*{Acknowledgments} 

We are grateful to A. Engel, L. Peliti, P. Eichelsbacher, W. Kirsch, 
T. Kriecherbauer and N. Kumar for useful discussions. This work was 
supported by DFG within 
SFB/TR 12 \textit{Symmetries and Universality in Mesoscopic Systems.}

\appendix

\section{Inter-record spacing for the shell model}
\l{appendix1}
Using the Stirling's formula $r ! \approx \sqrt{ 2 \pi r} (r/e)^{r}$ in  
the binomial coefficient ${N \choose r}$ for $r, N \gg 1$ with $r/N$ fixed and 
expanding ${N \choose r}$ about its maximum at $r=N/2$, we have 
\be
\frac{1}{2^{N}} \; {N \choose r} \approx \sqrt{\frac{2}{\pi N}} \; 
\mbox{exp} \left[ {-\frac{(N/2-r)}{N/2}}^{2} \right] \;\;. \l{cnr}
\ee
This expression can be used to show that
\bea
\frac{1}{2^{N}} \; \sum_{r=0}^{y}{N \choose r} & \approx & 
\frac{1}{2} \mbox{erfc}(\alpha) \l{sumcnr} \\
\frac{1}{2^{N}} \; \sum_{r=0}^{y} r {N \choose r} & \approx & \frac{N}{4} 
\left( \mbox{erfc}( \alpha) - \sqrt{\frac{2}{\pi N}} \mbox{exp} 
(- \alpha^{2}) \right) \;\;,\l{avgcnr}
\eea
by approximating the sum on the left hand side by an integral. In 
the above expressions, 
$\alpha = (N/2-y)/\sqrt{N/2}$ and 
$\mbox{erfc(x)} = (2/\sqrt{\pi}) \int_{x}^{\infty} dt \; e^{-t^{2}}$ is 
the complementary error function. One can derive (\ref{rj}) for the 
average location $\av{r_{j}}=\sum r_{j} \tilde{P}(r_{j})$ where 
$\tilde{P}(r_{j})$ is given by (\ref{recloc}) using 
Eqs.(\ref{cnr})-(\ref{avgcnr}) and replacing the sums by integrals. 

The integral $G(m)$ in Section~\ref{recordspace} can be estimated by saddle 
point method as follows. We have 
\be
G(m) = (-1)^{m} \int_{-\infty}^{\infty} dx \; 
\frac{e^{-2 x^{2}}}{\mbox{erfc}(x)} \; 
\frac{\left(\ln {\mbox{erfc(x)}}\right)^{m}- \left( \ln {2} \right)^{m}}{m!}
\approx  (-1)^{m} 
\int_{-\infty}^{\infty} dx \; \frac{e^{-{\cal{G}}(x)}}{m!} \;\;,
\ee
where ${\cal{G}}(x) \approx x^{2} - \ln x \; (2 m + 1+ x^{-2})$ for 
large $m$. Here we have used 
that $\mbox{erfc}(x) \approx e^{-x^{2}}/ (\sqrt{\pi} x) $ for $x \gg 1$. 
Expanding ${\cal{G}}(x)$ about its minimum $x_{0} \approx m- \ln \sqrt{m}$ 
up to $O(x-x_{0})^{2}$ and doing the Gaussian 
integral, we obtain $G(m) \approx \sqrt{ \pi m/2}$ for large $m$. 

\section{Distribution of the most populated sequence}
\l{appendix2}

Our goal is to compute the probability $P_t(k^\ast)$ that the 
sequence $k^{\ast}$ has 
the maximum population at time $t$ in the i.i.d. model. 
This distribution is given by 
\be  
P_t(k^\ast) = \int_{E_{\mathrm{min}}(k^\ast,t)}^{E_{\mathrm{max}}(k^\ast,t)} 
dE \;\; p_t^{(k^\ast)}(E)  \;\; 
\prod_{k \neq k^{\ast}} 
q_t^{(k)}(E) \;\;,
\ee 
where 
\be
p_t^{(k)}(E) = t^{-1} p[(E+k)/t] 
\ee
is the distribution of $E(k,t)$ obtained from the fitness distribution
$p(F)$ via the variable transformation (\ref{lineq}). The 
limits of the support of $p_t^{(k)}(E)$ are 
$E_{\mathrm{min}}(k,t) = -k+ F_{\mathrm{min}} t$
and $E_{\mathrm{max}}(k,t) = -k+ F_{\mathrm{max}} t $, and 
$q_t^{(k)}(E)= \int_{E_{\mathrm{min}}}^{E} dE \; p_t^{(k)}(E) $
is the corresponding cumulative distribution for $E > E_{\mathrm{min}}$ 
and zero otherwise. In the following, we show 
that the distribution $P_t(k^\ast)$ is of the scaling form 
$P_t(k^\ast) \approx t^{-1/z} \; \Phi (k^{\ast}/{t^{1/z}})$ 
for various choices of fitness distribution. 

(i) $\;$ \underline{$p(F)=e^{-F}$:} 
\bea
P_t(k^{\ast}) = \frac{1}{t} 
\int_{0}^{\infty} dx \; e^{-(x+k^{\ast})/t} 
\prod_{k \neq k^{\ast}} (1-e^{-(x+k)/t}) \approx \frac{1}{t} e^{-k^{\ast}/t}  \;\;,
\eea
where the last expression is obtained by exponentiating the product and 
evaluating the resulting sum as an integral for an infinite system.  
Thus in this case $z = 1$ and the scaling function is given by 
$\Phi_1(y) = e^{-y}$.

(ii) $\;$ \underline{$p(F)=(1+\nu) (1-F)^{\nu}$, $\nu \geq -1$:}
\bea
P_t(k^\ast) &=& \frac{1+\nu}{t} \int_{0}^{t-k^{\ast}} dx \; 
\left( 1- \frac{x+k^{\ast}}{t} \right)^{\nu} 
\prod_{k \neq k^{\ast}} 
\left[ 1-\left( 1- \frac{x+k}{t} \right)^{1+\nu} \right] \\
 & \approx &  \frac{1+\nu}{t} \int_{0}^{t-k^{\ast}} dx \; 
\left( 1- \frac{x+k^{\ast}}{t} \right)^{\nu} 
\mbox{exp} \left[ - \frac{t}{2+\nu} 
\left( 1- \frac{x}{t} \right)^{2+\nu} \right] \\
& \approx & \frac{1}{t^{1/z}} \; \Phi_{2} \left( y= \frac{k^{\ast}}{t^{1/z}} 
\right) \;\;,
\eea
where $z=(2+\nu)/(1+\nu)$ and the scaling function $\Phi_{2}(y)$ is given by
\be
\Phi_{2}(y) \approx  \frac{1+\nu}{(2+\nu)^{1/z}} 
\int_{y^{2+\nu}/(2+\nu)}^{\infty} dx \; 
e^{-x} x^{-1/z} \left( y+ ((2+\nu) x)^{1/(2+\nu)} \right)^{\nu} \;\;.
\ee

(iii) $\;$ \underline{$p(F)= \delta \; F^{-1-\delta}$, $\delta > 1$:}
\bea
P_t(k^\ast) &=& \frac{\delta}{t} \int_{t}^{\infty} dx \; 
\left( \frac{t}{x+k^{\ast}} \right)^{1+\delta} 
\prod_{k \neq k^{\ast}} \left[ 1- \left( \frac{t}{x+k} \right)^\delta 
\right] \\
& \approx & \frac{1}{t^{1/z}} \; \Phi_{3} \left( y= \frac{k^{\ast}}{t^{1/z}} 
\right) \;\;,
\eea
where $z=(\delta-1)/\delta$ and the scaling function $\Phi_{3}(y)$ 
is given by 
\be
\l{f3}
\Phi_{3}(y) \approx \delta (\delta -1)^{1/(\delta -1)}
\int_{0}^{\infty} dx 
\frac{e^{-x} x^{1/(\delta-1)}}{(1+((\delta-1) x)^{1/(\delta-1)}y)^{1+\delta}} 
\;\;.
\ee


\begin{figure}
\begin{center}
\psfig{figure=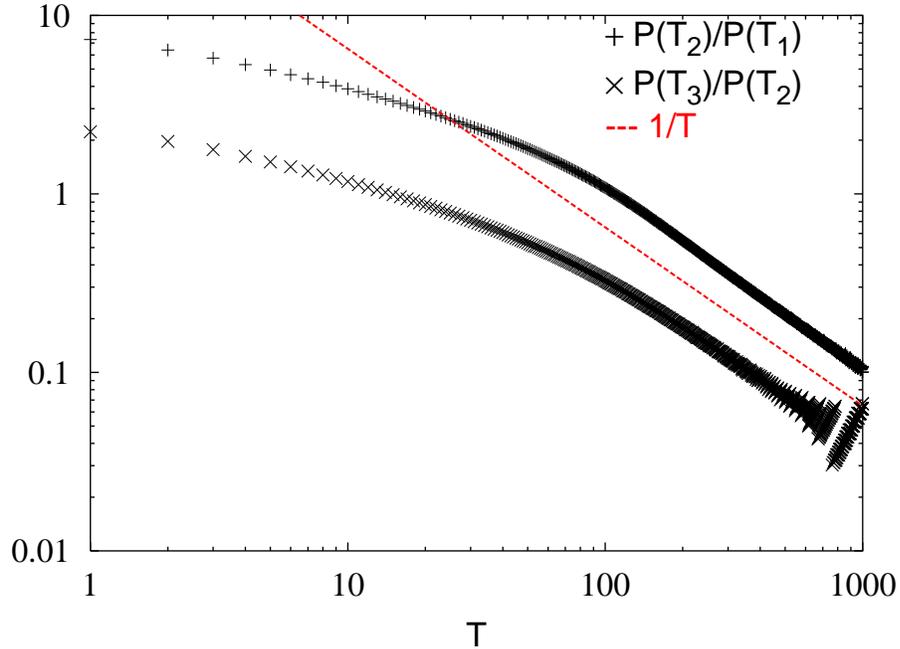,width=9cm,angle=270}
\caption{Log-log plot for the ratios $P(T_{2})/P(T_{1})$ and 
$P(T_{3})/P(T_{2})$ of evolution time distributions
for the i.i.d. model with $p(F)=e^{-F}$. The ratios decay as 
$1/T$ for large $T$, as the line with 
slope $-1$ indicates.}
\label{evol}
\end{center}
\end{figure}

\begin{figure}
\begin{center}
\psfig{figure=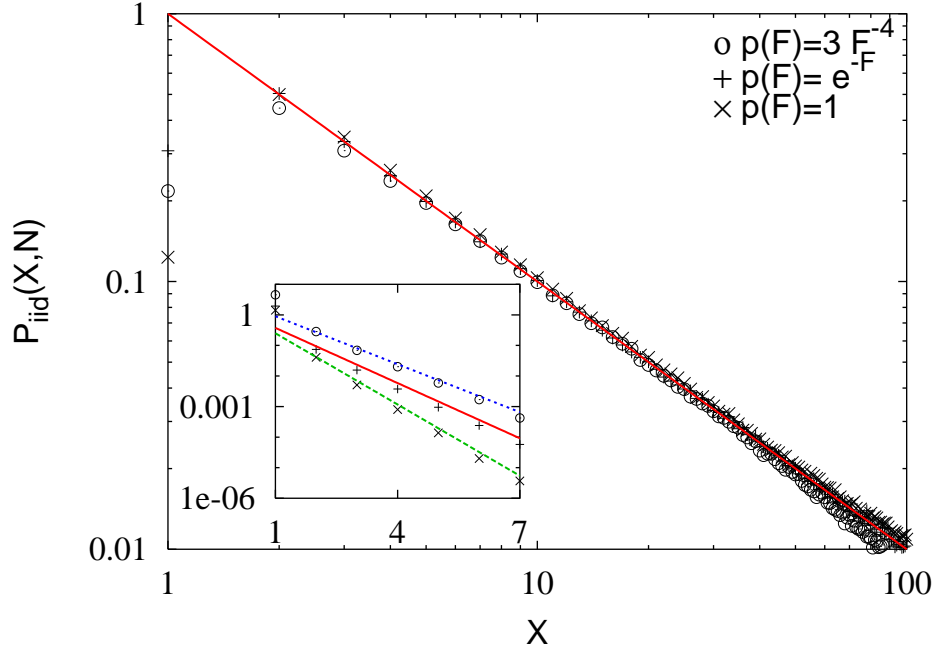,width=9cm,angle=270}
\caption{Distribution 
$P_{\mathrm{iid}}(X,N)=P_{\mathrm{iid}}(t=X,N)+P_{\mathrm{iid}}(k=X,N)$ vs. 
$X$ for various distributions, illustrating the 
duality between the jump processes along the $k$- and $t$-axes. The 
solid line has slope $-1$. Inset: Semi-log plot for the inter-jump spacing 
$\Delta_{\mathrm{iid}}^{\prime}(j)$ along the time axis as a function of 
$j$ for the i.i.d. model. The slope $(1-\beta)/2$ is consistent with the 
conjecture (\ref{cnjctre}).}
\label{sum}
\end{center}
\end{figure}

\begin{figure}
\begin{center}
\psfig{figure=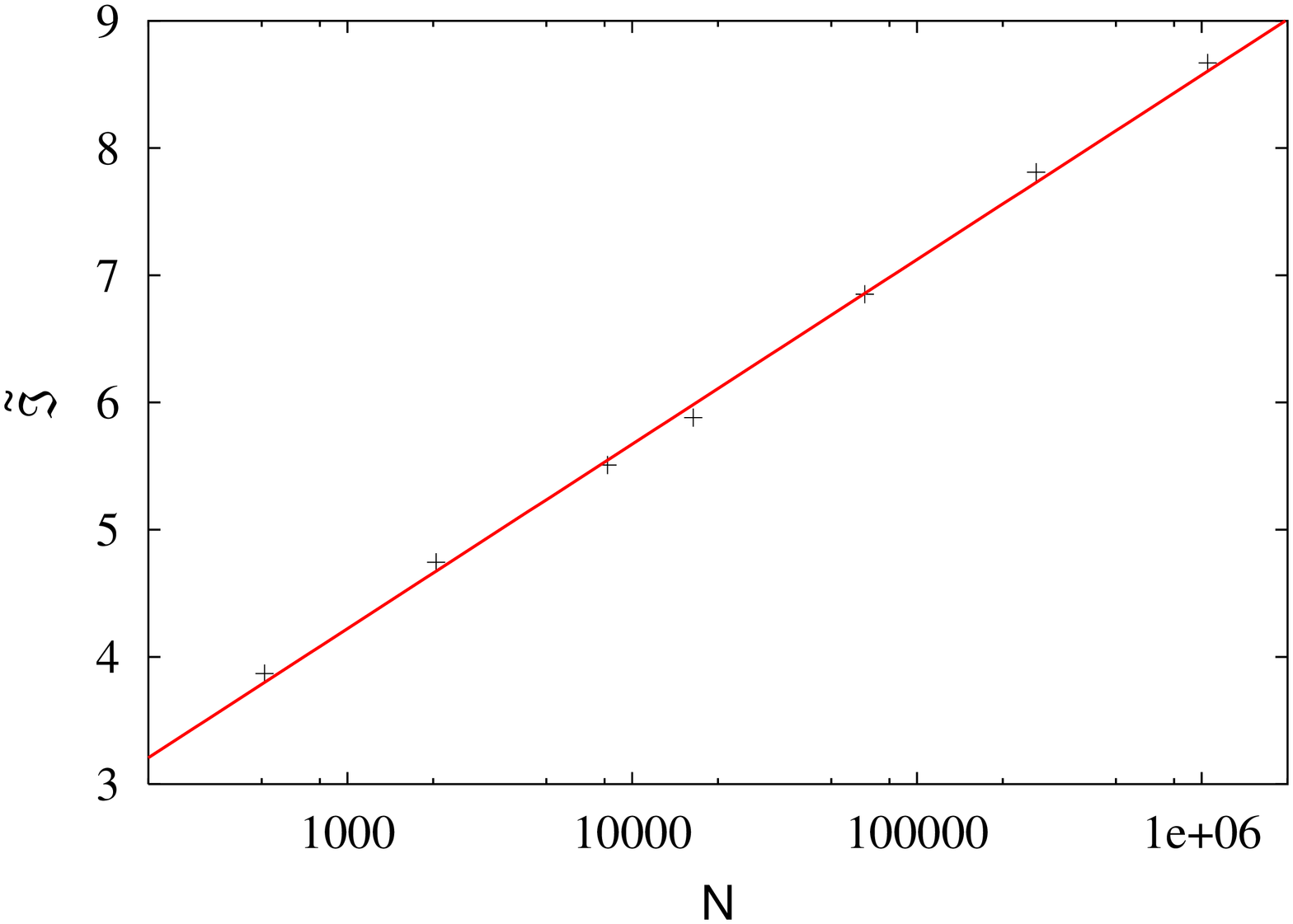,width=9cm,angle=270}
\caption{Average number $\tilde {\cal{J}}$ of jumps 
vs. $N$ for the record time model defined by (\ref{mfmodel}). The solid line 
$0.63 \ln N-0.13$ is the best fit.}
\label{mftjump}
\end{center}
\end{figure}



\begin{thebibliography}{999}
\bibitem{remark} This scaling form is not true for 
power law distributed 
fitness with $1 < \delta \leq 2$, because the scaling function $\Phi$ in
(\ref{Ptscale}) does not possess a first moment in this case (see (\ref{f3})). 
Nevertheless, the time scale 
$T^{\ast} \sim N^{z}$ with $z$ given in (\ref{1dz}) for all $\delta > 1$.
\end{thebibliography}
\end{document}